\documentclass[sigconf]{acmart}
\usepackage{amsmath,amsfonts}
\usepackage{fnpct}
\usepackage{algorithmic}
\usepackage{graphicx}
\usepackage{textcomp}
\usepackage{xcolor}
\usepackage{tcolorbox} 
\usepackage{balance}
\usepackage{array} 
\usepackage{multirow}
\usepackage{pifont} 

\usepackage{fancyhdr}
\usepackage{tikz}
\usepackage{makecell}
\usepackage{amsmath} 

\usetikzlibrary{shapes.geometric, arrows.meta, positioning, calc}

\AtBeginDocument{%
  }

\copyrightyear{2026}
\acmYear{2026}
\setcopyright{cc}
\setcctype{by}
\acmConference[CODASPY '26]{Proceedings of the Sixteenth ACM Conference on Data and Application Security and Privacy}{June 23--25, 2026}{Frankfurt am Main, Germany}
\acmBooktitle{Proceedings of the Sixteenth ACM Conference on Data and Application Security and Privacy (CODASPY '26), June 23--25, 2026, Frankfurt am Main, Germany}
\acmDOI{10.1145/3800506.3803490}
\acmISBN{979-8-4007-2562-3/2026/06}

\begin{document}

\title[Reality Check on SBOM Vulnerability Management]{A Reality Check on SBOM-based Vulnerability Management: An Empirical Study and A Path Forward}

\author{Li Zhou}
\email{li.zhou@kaust.edu.sa}
\orcid{0009-0007-1222-6815}
\affiliation{%
  \institution{King Abdullah University of Science and Technology}
  \city{Thuwal}
  \country{Saudi Arabia}
}

\author{Marc Dacier}
\email{marc.dacier@kaust.edu.sa}
\orcid{0000-0003-3206-2030}
\affiliation{%
  \institution{King Abdullah University of Science and Technology}
  \city{Thuwal}
  \country{Saudi Arabia}
}

\author{Charalambos Konstantinou}
\email{charalambos.konstantinou@kaust.edu.sa}
\orcid{0000-0002-3825-3930}
\affiliation{%
  \institution{King Abdullah University of Science and Technology}
  \city{Thuwal}
  \country{Saudi Arabia}
}
\renewcommand{\shortauthors}{Li Zhou, Marc Dacier, \& Charalambos Konstantinou}

\begin{abstract}
The Software Bill of Materials (SBOM) is a critical tool for securing the software supply chain (SSC), but its practical utility is undermined by inaccuracies in both its generation and its application in vulnerability scanning. This paper presents a large-scale empirical study on 2,414 open-source repositories to address these issues from a practical standpoint. First, we demonstrate that using lock files with strong package managers enables the generation of accurate and consistent SBOMs, establishing a reliable foundation for security analysis. Using this high-fidelity foundation, however, we expose a more fundamental flaw in practice: downstream vulnerability scanners produce a staggering 92.0\% false positive rate in our case study. We pinpoint the primary cause as the flagging of vulnerabilities within unreachable code. We then demonstrate that function call analysis can effectively prune 61.9\% of these false alarms. Our work validates a practical, two-stage approach for SSC security: first, generate an accurate SBOM using lock files and strong package managers, and second, enrich it with function call analysis to produce actionable, low-noise vulnerability reports that alleviate developers' alert fatigue.

\end{abstract}



\maketitle

\begin{table*}[!t]
\centering
\setlength{\abovecaptionskip}{4pt}
\setlength{\belowcaptionskip}{0pt}
\caption{Summary of related works on SBOM tools and SBOM-based vulnerability scanners with their associated languages/platforms.}
\label{tab:literature}
\footnotesize
\resizebox{1.8\columnwidth}{!}{%
\begin{tabular}{l>{\raggedright\arraybackslash}p{9cm}>{\raggedright\arraybackslash}p{4cm}}
\toprule
\textbf{Research} & \textbf{SBOM Tools \& SBOM-based Vulnerability Scanners} & \textbf{Languages/Platforms}  \\
\midrule
\citet{yu2024correctness} & \texttt{Syft}~\cite{syft}, \texttt{Trivy}~\cite{trivy}, \texttt{GitHub-DG}~\cite{github_dependency_graph}, \texttt{Sbom-tool}~\cite{microsoft_sbom_tool} & Python, Java, JavaScript, Go, .NET, PHP, Ruby, Rust, Swift \\
\citet{rabbi2024sbom} & \texttt{Cdx-node-module}~\cite{cyclonedx_node_module}, \texttt{Ort}~\cite{oss_review_toolkit}, \texttt{Syft}~\cite{syft}, \texttt{Cdxgen}~\cite{cyclonedx_cdxgen} & JavaScript \\
\citet{o2023impacts} & \texttt{CVE-bin-tool}~\cite{intel_cve_bin_tool}, \texttt{Grype}~\cite{anchore_grype}, \texttt{Trivy}~\cite{trivy} & Docker, Binary \\
\citet{benedetti2024impact} & \texttt{GitHub-DG}~\cite{github_dependency_graph}, \texttt{Ort}~\cite{oss_review_toolkit}, \texttt{Syft}~\cite{syft}, \texttt{Trivy}~\cite{trivy}, \texttt{Cdxgen}~\cite{cyclonedx_cdxgen} & Python \\
\citet{balliu2023challenges} & \texttt{Build-info-go}~\cite{jfrog_build_info_go}, \texttt{Depscan}~\cite{owasp_dep_scan}, \texttt{OpenRewrite}~\cite{openrewrite}, \texttt{Cdx-maven-plugin}~\cite{cyclonedx_maven_plugin}, \texttt{Cdxgen}~\cite{cyclonedx_cdxgen}, \texttt{Jbom}~\cite{eclipse_jbom} & Java \\
\citet{o2024assessing} & \texttt{Grype}~\cite{anchore_grype}, \texttt{Trivy}~\cite{trivy} & Docker \\
\citet{reinhold2023new} & \texttt{CVE-bin-tool}~\cite{intel_cve_bin_tool}, \texttt{CWE-checker}~\cite{fkie_cwe_checker} & Binary \\
\citet{cofano2024sbom} & \texttt{Grype}~\cite{anchore_grype}, \texttt{Ort}~\cite{oss_review_toolkit}, \texttt{Trivy}~\cite{trivy}, \texttt{Cdxgen}~\cite{cyclonedx_cdxgen} & Python \\
\bottomrule
\end{tabular}
}
\end{table*}

\section{Introduction}

Software often relies on third-party libraries as dependencies~\cite{mockus2007large, ellison2010evaluating}. These libraries, in turn, have their own dependencies, referred to as transitive dependencies. These hierarchical dependencies form a software supply chain~(SSC)~\cite{ellison2010evaluating}. Due to the transitive and often opaque nature of third-party dependency management, SSC-based attacks pose significant challenges to software security~\cite{alkhadra2021solar, lakshmanan2023pytorch, lins2024critical}. To mitigate such attacks, the United States National Telecommunications and Information Administration~(NTIA) proposed the utilization of a software bill of materials~(SBOM)~\cite{beg2023supplychain, ntia_sbom}. An SBOM records all direct and transitive dependencies required during the software build process to trace the SSC. In practice, software developers leverage SBOMs to identify potential security risks, thereby enhancing the security of the SSC~\cite{o2023impacts, benedetti2024impact}. SBOMs provide insights into the software dependencies. Therefore, it is critical to ensure their accuracy and quality for subsequent analyses.

Recent studies have highlighted the limitations of existing SBOM generators in producing accurate and reliable SBOMs~\cite{yu2024correctness, o2023impacts, o2024assessing, rabbi2024sbom, zhao2024covsbom}. For instance, \citet{yu2024correctness} pointed out that, given the same input files, different SBOM generators often produce SBOMs with inconsistent and inaccurate results. Similarly, \citet{o2023impacts}~revealed that discrepancies in SBOM outputs from various generators significantly impact the effectiveness of downstream vulnerability analyses. Moreover, \citet{balliu2023challenges} demonstrated that even two official SBOM generators provided by CycloneDX~\cite{cyclonedx}, the founder of a popular SBOM standard with the same name, produce different results with considerable variations. 

Despite the industrial consensus on using lock files for reproducibility~\cite{GitLab_SBOM_Doc, Snyk_SBOM_CLI_Doc, Ostorlab_SBOM_Tutorial}, a concerning methodological disconnect persists in the academic community. As shown in Table~\ref{tab:literature}, numerous recent studies~\cite{yu2024correctness, o2023impacts, o2024assessing, rabbi2024sbom, zhao2024covsbom} continue to rely on project manifest files as the primary input for vulnerability analysis, inevitably yielding erroneous outcomes.  \citet{yu2024correctness} critically exposed the inconsistencies inherent in SBOM generated from project files, which developers use to describe direct dependencies, and advocated for the adoption of lock files. Unfortunately, their claim lacked large-scale empirical experimental validation. In this work, we close this gap by rigorously validating their proposition. We show  that inconsistencies observed in prior work are not inherent to the SBOM tools themselves but, rather, are due to the ambiguous nature of the input provided to them. We demonstrate that adhering to lock files and strong package managers (PMs), which are capable of managing direct and transitive dependencies (further details on PMs and their categorization are provided in Section~\ref{ss:pms}), is not merely a technical preference but a prerequisite for correctness. Consequently, we argue that future research must standardize on lock files as the definitive input for SBOM generation to ensure accuracy and completeness. We hope that our work will close this debate once and for all. 

By establishing a method to generate accurate SBOMs, we create a reliable baseline to empirically investigate and quantify a more profound flaw in the downstream vulnerability pipeline. Existing SBOM-based vulnerability scanners often suffer from a high rate of false positives~\cite{zhao2024covsbom, cofano2024sbom}. These tools operate by matching the versions of third-party components listed in an SBOM against a vulnerability database, which maintains records of known vulnerabilities and their affected version ranges. The primary source of these inaccuracies lies in this version-only matching approach. Although a component's version may be listed as vulnerable, the specific code containing the vulnerability might not be invoked by the application. Consequently, developers are presented with reports containing numrous irrelevant alerts, which creates a significant verification burden and leads to alerts fatigue, a well-documented phenomenon where users become desensitized and start to ignore frequent or repetitive warnings~\cite{AHRQ_Alert_Fatigue}, increasing the risk that genuine, critical vulnerabilities are overlooked. 

We conduct a large-scale experiment on 2,414 open-source repositories across four programming languages to address these challenges systematically. First, we establish that using lock files as input enables SBOM generators to produce consistent and accurate results, creating a high-fidelity foundation for analysis. Second, using this foundation, we quantify the end-to-end performance of the SBOM-based vulnerability management pipeline, revealing a staggering false positive rate of 92.0\% in the manually validated dataset and pinpointing unreachable code as the primary cause. Finally, we validate that function call analysis can effectively prune a significant number of these false positives by 61.9\%, paving the way for a more practical vulnerability management strategy that significantly reduces false positives and alleviates developers' alert fatigue.

Our contributions are summarized as follows:
\begin{itemize}
    \item We propose to use the lock file instead of the previously used project file as input to SBOM generators, and establish the importance of doing so in order to produce accurate and complete SBOMs. Our large-scale experiments demonstrate that SBOM generators can generate accurate and consistent SBOMs with lock files. 
    \item Using these accurate SBOMs, we conduct a large-scale analysis of downstream vulnerability reports. We quantify the significant false positive rates of existing scanners and pinpoint unreachable code as a primary cause of these false alarms.
    \item We propose a best-practice pipeline for generating more reliable vulnerability reports. We validate this approach through function call analysis, demonstrating its effectiveness in significantly reducing false positives and guiding developers toward more effective security practices. We open-source our code on GitHub\footnote{\label{note1}https://github.com/damaoooo/validation}\footnote{\label{note4}https://github.com/damaoooo/SBOMVerifierGoPoC} with the corresponding SBOM and vulnerability verification artifacts.
\end{itemize}

The remainder of this paper is organized as follows: Section~\ref{sec:background} provides background for this study. We detail our motivation for this research in Section~\ref{sec:motivation}. Section~\ref{sec:methodology} details our methodology and evaluation setup. Our experimental results are demonstrated in Section~\ref{sec:evaluation}. In Section~\ref{sec:discussion}, we examine the reasons behind the inconsistencies in SBOM generator outputs and propose best practices for enhancing SSC security. We conclude our paper and explore future directions in Section~\ref{sec:conclusion}.

\begin{table*}[!t]
\centering
\setlength{\tabcolsep}{3pt}
\renewcommand{\arraystretch}{0.95}
\scriptsize
\caption{Comparison of common package managers across programming languages.}
\label{tab:package_manager_comparison}
\resizebox{2\columnwidth}{!}{%
\begin{tabular}{llcccc>{\raggedright\arraybackslash}p{4cm}}
\toprule
\textbf{Language} & \textbf{Package Manager} & \textbf{Strong PM} & \textbf{Lock File} & \textbf{Dry-Run Support} & \textbf{SBOM} & \textbf{Remarks} \\
\midrule
\textbf{Python} & Poetry & \ding{51} & \texttt{poetry.lock} & \ding{51} & \ding{51} & Poetry can also be used to cooperate with other PMs. \\
& Pip & \ding{55} & \ding{55} & \ding{55} & \ding{51} & Standard \texttt{requirements.txt} files can be used for other Strong PMs. \\
\textbf{Golang} & Go Modules & \ding{51} & \texttt{go.sum} & \ding{51} & \ding{55} & Existing SBOM generators accept the \texttt{go.sum} files only for early Golang versions. \\
\textbf{Rust} & Cargo & \ding{51} & \texttt{Cargo.lock} & \ding{51} & \ding{51} & - \\
\textbf{Java} & Gradle & \ding{51} & \texttt{gradle.lockfile} & \ding{55} & \ding{51} & Gradle does not produce a lock file by command line.
\\
\textbf{JavaScript} & NPM & \ding{51} & \texttt{package-lock.json} & \ding{51} & \ding{55} & - \\
& Yarn & \ding{51} & \texttt{yarn.lock} & \ding{55} & \ding{55} & - \\
\textbf{Ruby} & Bundler & \ding{51} & \texttt{gemfile.lock} & \ding{51} & \ding{51} & Gem bundle supports listing cross-platform dependencies in one Gemfile. \\
\textbf{PHP} & Composer & \ding{51} & \texttt{Composer.lock} & \ding{55} & \ding{51} & - \\
\bottomrule
\end{tabular}
}
\end{table*}

\section{Background}\label{sec:background}
\subsection{Software Bill of Materials (SBOM)}

\subsubsection{SBOM Definition}

The NTIA defines an SBOM as a formal record documenting the components and their supply chain relationships used in software development~\cite{ntia_sbom}. It is a detailed inventory of the software components used in a project. It enables effective tracking and management of dependencies, helping to identify vulnerabilities and mitigate security risks. Typically, it includes information not only about component \texttt{names} and \texttt{versions} but also additional information, like \texttt{manufacturer}, \texttt{description}, \texttt{licenses}, \texttt{package website}, \texttt{checksum}, and other metadata~\cite{muiri2019framing}. Accurately identifying software components is vital for retrieving additional related information. Therefore, in this study, as a foundational step, we focus on verifying the correctness of the \texttt{name} and \texttt{version} of each component, aligning with previous research~\cite{yu2024correctness, o2024assessing,balliu2023challenges}. There are several established SBOM formats and standards~\cite{spdx_website,cyclonedx, waltermire2016guidelines}. Among these, CycloneDX~\cite{cyclonedx} and SPDX~\cite{spdx_website} are two of the most widely adopted formats, offering detailed package information~\cite{sehgal2023taxonomy}. Hence, in this study, we select CycloneDX and SPDX for our evaluation~\cite{cyclonedx, spdx_website}.

\subsubsection{SBOM Applications}

By listing the necessary software components, in theory, SBOMs should enable users to identify vulnerabilities and assess the security posture of software binaries without having access to their source code~\cite{benedetti2024impact}. Through a detailed inventory of software components and their versions, the authors in~\cite{o2023impacts} proposed referencing this information with vulnerability databases to detect vulnerable components and evaluate potential risks. Furthermore, SBOMs facilitate license management by providing information about the licenses associated with each software component. This ensures compliance with legal requirements and helps organizations identify potential licensing conflicts~\cite{dalia2024sbom}. Unfortunately, the poor quality of the SBOM tools output renders these tasks practically impossible, as explained in the previously mentioned references~\cite{yu2024correctness, o2023impacts, o2024assessing, balliu2023challenges}.

\subsection{Package Managers (PMs)}\label{ss:pms}

PMs are widely used in software engineering to manage third-party dependencies~\cite{muhammad2019taxonomy}. PMs typically use a so-called ``project file'' that lists the required third-party libraries, to handle transitive dependencies automatically, to verify whether these dependencies are available, and to install the third-party packages based on the resolved dependencies~\cite{spinellis2012package}. Nowadays, mainstream programming languages are equipped with their own PMs~\cite{decan2019empirical}. 

It is important to note that third-party libraries often have their own dependencies. Depending on how these transitive dependencies are managed, we categorize PMs into \textbf{strong PMs} and \textbf{weak PMs}. Strong PMs explicitly resolve all dependencies, both direct and transitive, and generate a lock file that records the exact versions of each dependency used. In contrast, weak PMs do not explicitly resolve transitive dependencies in a lock file; instead, they only retrieve transitive dependencies during the package installation process. The work in \cite{yu2024correctness} highlighted that missing transitive dependencies present a significant challenge for SBOM tools in generating accurate SBOMs. In this study, we demonstrate that this can be effectively bypassed by directly utilizing the lock files generated by strong PMs.

In essence, project files and lock files serve distinct roles in dependency management. Project files serve as declarative manifests, listing only direct dependencies and their acceptable version ranges while omitting transitive dependencies. Lock files, on the other hand, are automatically generated and offer a snapshot of the dependency tree at the time of installing all dependencies, detailing all the dependencies (including transitive and indirect ones) along with the exact version used. In other words, the project file specifies how all dependencies can be resolved to build the software (e.g., version \texttt{x} with \texttt{1.2.1 < x < 1.5.3}), whereas the lock file says which specific version has been used to build a given software. This distinction is pivotal for generating accurate SBOMs, as lock files ensure precise tracking of dependencies and their versions.


\subsection{SBOM-based Vulnerability Scanners}

Typically, an SBOM-based vulnerability scanner accepts a standard SBOM report as input. Then it queries the vulnerability databases to generate a corresponding vulnerability report. For each identified vulnerability, the report provides details from sources such as the Common Vulnerabilities and Exposures (CVE) database or GitHub Security Advisories (GHSA), along with specifics on how the vulnerability was matched~\cite{o2024assessing}. In practice, SBOM-based scanners are integrated directly into the Continuous Integration and Continuous Delivery/Deployment (CI/CD) pipeline~\cite{shahin2017continuous}. They act as automated quality gates within platforms such as GitHub Actions~\cite{github-actions} or Jenkins~\cite{jenkins-ci}, scanning container images and dependencies on every code commit~\cite{gitlab-container-scanning-docs, gocodeo-using-grype-ci-cd}. This approach provides developers with immediate feedback, allowing them to fix vulnerabilities before deployment and enforcing security policies by automatically failing builds that contain high-severity issues. In this study, we select the popular scanners \texttt{Grype}~\cite{anchore_grype} and \texttt{Trivy}~\cite{trivy} for our evaluation.

The Vulnerability Exploitability eXchange (VEX)~\cite{ntia_vex_2021} has been proposed as a companion of SBOM that aims to contextualize scanner findings. A VEX statement can, for example, declare a vulnerability as \texttt{not\_affected} if the vulnerable code is unreachable. However, VEX is merely a communication format rather than an analysis engine. It does not verify the vulnerability report, but instead, transfers the significant burden from the software consumer to the producer responsible for generating the VEX advisory. Thus, the underlying challenge of numerous false positives remains unsolved.

\subsection{Research Landscape and Positioning}
\label{sec:related_work}
As summarized in Table~\ref{tab:literature}, recent studies evaluate SBOM generators and vulnerability scanners. We categorize them into three primary dimensions to highlight our novel contributions.

\textbf{Inconsistencies in SBOM Generation:} 
Several studies highlight the inconsistent or incomplete SBOM outputs across SBOM tools~\cite{yu2024correctness, rabbi2024sbom, benedetti2024impact, cofano2024sbom, balliu2023challenges},  Most of these evaluations rely on project manifest files (e.g., \texttt{requirements.txt}), which do not deterministically encode transitive dependencies. In contrast, we use lock files as the evaluation baseline and show that this choice substantially improves reproducibility and consistency.

\textbf{Limitations of Current Vulnerability Scanners:} 
Existing studies~\cite{o2023impacts, o2024assessing, benedetti2024impact} leverage SBOM-based scanners to assess software security risks, but their conclusions are drawn without analyzing actual vulnerability reachability. However, current scanners operate purely at a coarse-grained, package level via version matching, which can inevitably overapproximate and inflate false alarms. Our results quantify this issue with a 92.0\% false positive rate in our case study. This limitation also appears in binary-focused tooling such as \texttt{CVE-bin-tool}~\cite{intel_cve_bin_tool, reinhold2023new}, where vulnerability granularity remains coarse.

\textbf{Vulnerability Reachability Analysis:} 
To alleviate alert fatigue, efforts like CovSBOM~\cite{zhao2024covsbom} integrated code coverage into \texttt{Java} SBOMs. Inspired from this, our work provides the first multi-language empirical vulnerability check. We validate that function-call analysis prunes a considerable portion of false alarms. We also identify a key practical bottleneck: mainstream vulnerability sources (e.g., NVD, GHSA) often lack structured symbol-level metadata needed for end-to-end automation.

\section{Motivation}\label{sec:motivation}

This section outlines the motivation for this research. Existing literature, with findings detailed in Table~\ref{tab:literature}, reveals that current tools often fail to generate accurate and consistent SBOMs. Therefore, this research aims to fix this deficiency by proposing reliable approaches for SBOM generation. As suggested by the NTIA, SBOMs are crucial for enhancing SSC security and can facilitate various downstream tasks. One prominent example is the detection of vulnerabilities resulting from third-party library reuse, with several studies proposing the use of SBOMs for this purpose~\cite{o2023impacts, anchore_grype, benedetti2024impact}. Consequently, the inability of current tools to produce reliable SBOMs directly undermines the effectiveness of these critical applications, rendering downstream analyses unreliable. 

Specifically, \citet{yu2024correctness} systematically evaluated SBOM tools and concluded that current tools often fail to generate consistent and accurate SBOMs. Based on the findings of~\cite{yu2024correctness}, we conducted an in-depth analysis of two tools, \texttt{Syft}~\cite{syft} and \texttt{Trivy}~\cite{trivy}. We discovered that inconsistencies in generated SBOMs frequently stem from the reliance on project files. Certain transitive dependencies and overly broad version constraints specifications like \texttt{package <= version} commonly used in Python, for instance, may not be accurately recognized, leading to inaccuracies in the resulting SBOMs. Building on our analysis of those tools, we observe that lock files, particularly those generated by strong PMs based on project files, faithfully record the exact versions of all installed dependencies, both direct and indirect. This level of complete and precise information is exactly what is required to generate accurate SBOMs, offering a reliable foundation compared to relying only on project files. Therefore, our work proposes to use these strong PMs in conjunction with their generated lockfiles to produce accurate and consistent SBOMs, thereby supporting the reliability of downstream tasks.

However, as we developed this reliable method, a more fundamental and unsettling question emerged. The community has operated under the implicit assumption that an accurate SBOM is the principal missing piece for effective vulnerability management. We are motivated to challenge this assumption. We hypothesized that even with a perfect, ground-truth SBOM, the coarse, package-level analysis performed by downstream scanners might still produce unreliable results. This hypothesis formed the second, more profound motivation for our research: to empirically test the validity of the entire end-to-end, SBOM-based vulnerability scanning pipeline, an area largely taken for granted by practitioners.

Therefore, the motivation for this paper is twofold. First, we aim to provide a definitive, practical solution to the known SBOM accuracy problem by prioritizing a lock-file-based approach. Second, we leverage these newly accurate SBOMs to conduct the first large-scale empirical reality check on the end-to-end vulnerability reporting pipeline. This two-stage investigation allows us not only to fix a known flaw in SBOM generation but also to expose a more critical limitation in their downstream application, thereby providing crucial insights for both researchers and practitioners.

\section{Methodology}\label{sec:methodology}

In this section, we outline the methodology employed to evaluate SBOM generators and vulnerability scanners. We introduce the selected SBOM generators, PMs, and vulnerability scanners, describe the process of benchmark data collection, and define the performance metrics utilized in the evaluation. This detailed setup provides a framework for evaluating the proposed methods and the performance of SBOM tools and vulnerability scanners.

\begin{figure*}[tb]
    \centering
    \includegraphics[width=0.8\linewidth]{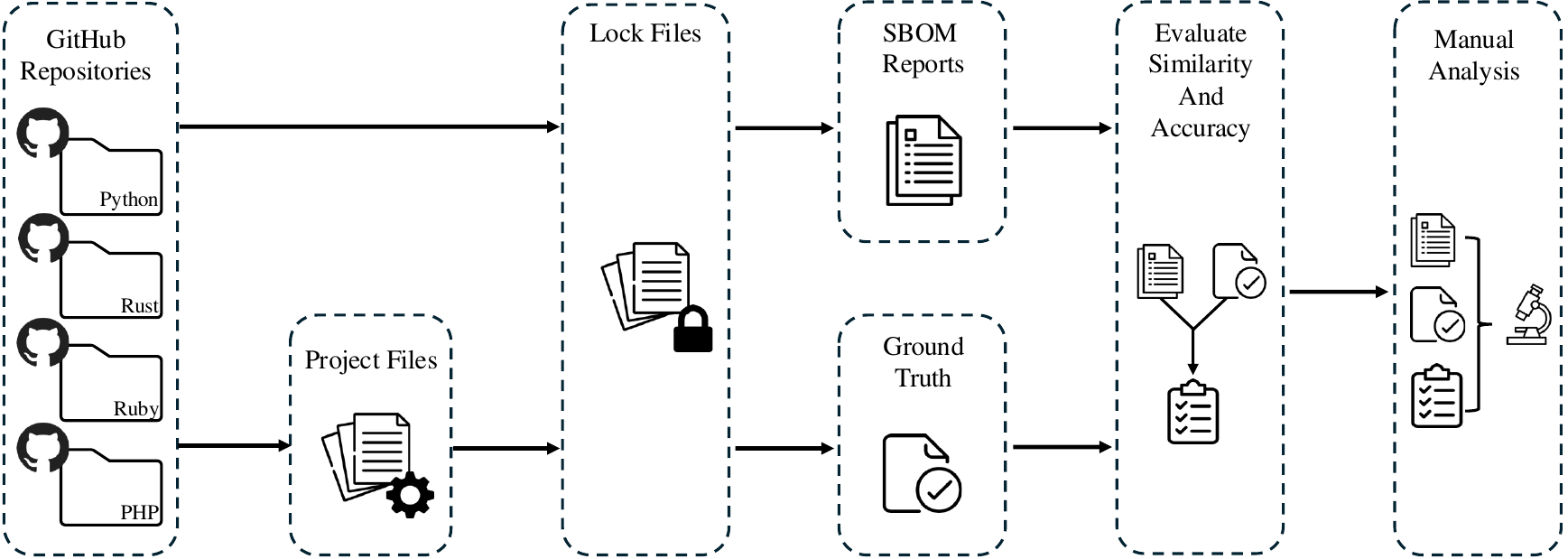}
    \caption{An overview of SBOM verification methodology.}
    \Description{A workflow diagram of the SBOM verification pipeline, from repository collection and lock-file processing to SBOM generation, comparison with ground truth, and metric evaluation.}
    \label{fig:methodology}
\end{figure*}

\begin{figure*}[tb]
    \centering
    \includegraphics[width=0.81\linewidth]{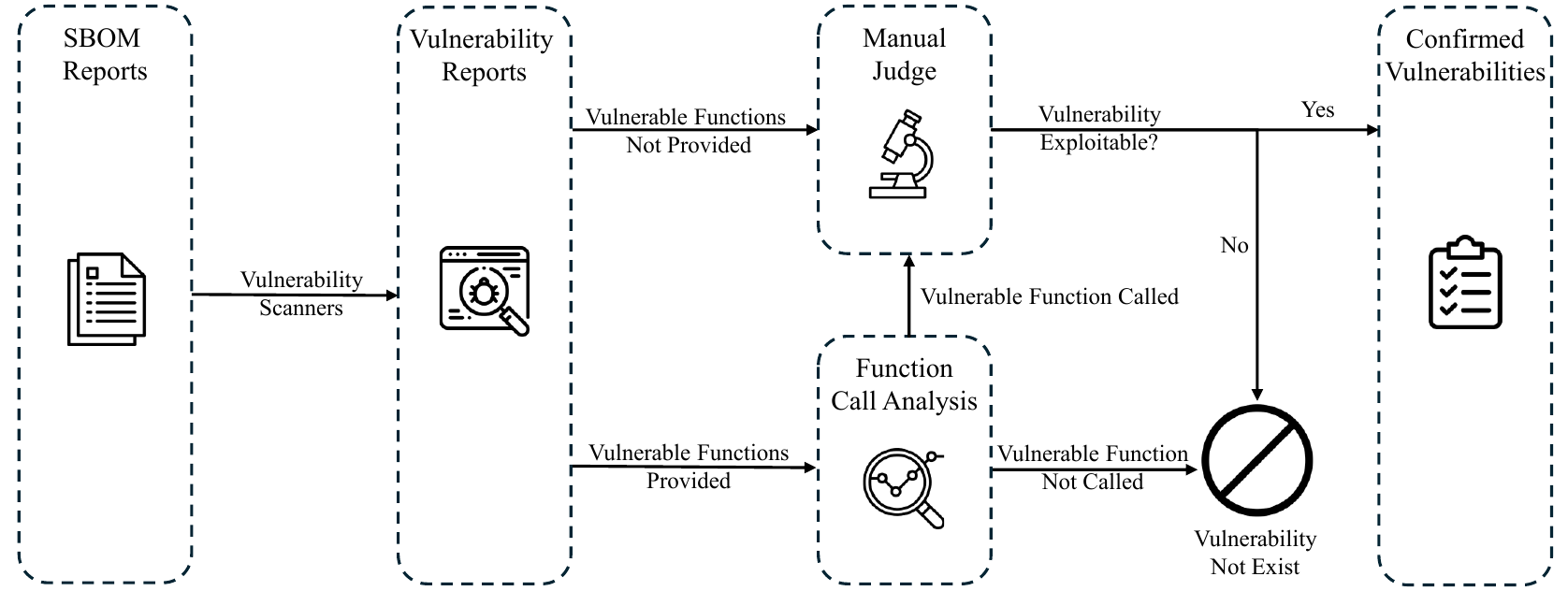}
    \caption{An overview of vulnerability verification methodology.}
    \Description{A workflow diagram of the vulnerability verification pipeline, showing SBOM-based reporting, manual or automated reachability checking, and filtering of false positive alerts.}
    \label{fig:vulnerablity}
\end{figure*}

\subsection{Overview}

Fig.~\ref{fig:methodology} and Fig.~\ref{fig:vulnerablity} illustrate the workflow of this study. We begin by collecting popular code repositories from GitHub and extracting their project files and lock files. For repositories that lack lock files, we use strong PMs corresponding to each programming language to generate a lock file from each project file. These newly generated lock files, along with the existing ones, serve as input for generating SBOM reports. We then compare the generated SBOMs against the ground truth data, which is extracted by parsing the lock files, to calculate Jaccard similarity and accuracy. For vulnerability detection, we first generate vulnerability reports from SBOMs generated by lock files for each repository. We then manually verify the existence of each vulnerability. Finally, we conduct a detailed analysis of the results to provide the insights discussed in Section~\ref{sec:discussion}.

\subsection{SBOM Generators and PMs}
In this work, we evaluate two popular SBOM generators: \texttt{Trivy}~\cite{trivy} and \texttt{Syft}~\cite{syft}. \texttt{Trivy} and \texttt{Syft} are open-source tools with extensive community support. Both tools support the standard CycloneDX format and support a wide range of programming languages and PM files. We select both tools to replicate consistently the method used in~\cite{yu2024correctness}, and to the best of our knowledge, those two are among the most widely used SBOM generators~\cite{yu2024correctness,benedetti2024impact,o2023impacts,rabbi2024sbom}.
However, recent studies highlight that the two tools demonstrate distinct approaches for dependency management and for the resolution of transitive dependencies. Consequently, their outputs can vary significantly~\cite{yu2024correctness, o2023impacts, o2024assessing}. In this study, we employ these two generators to demonstrate that, with the appropriate input, both \texttt{Trivy} and \texttt{Syft} are capable of generating identical, accurate, and complete SBOMs.

For commonly used programming languages, we list their respective common PMs as illustrated in Table~\ref{tab:package_manager_comparison}. 
The \textit{Lock File} column specifies the names of the generated lock files, while the \textit{Dry-Run Support} column indicates whether the lock file can be generated without having to install all dependencies locally, thereby conserving computational resources. The \textit{SBOM} column denotes whether the lock file is supported by selected SBOM generators. For instance, in Rust, which employs a strong PM, the project configuration file is \texttt{Cargo.toml}, and the corresponding lock file is \texttt{Cargo.lock}, which is recognized by SBOM tools. Using the same \texttt{Cargo.lock}, hence the software environment can be reliably replicated across different machines. 
We focus on strong PMs that generate lock files that are compatible with SBOM tools. Accordingly, we select \texttt{Poetry} for Python, \texttt{Cargo} for Rust, \texttt{Bundler} for Ruby, and \texttt{Composer} for PHP as the targets of our evaluation. 
It is worth mentioning that, although both \texttt{Trivy} and \texttt{Syft} claim to support \texttt{package-lock.json}, in our experiment, they produce empty SBOMs for this file. Therefore, we consider \texttt{package-lock.json} as unsupported by these tools~\footnote{We conducted our experiment using \texttt{Trivy} version 0.66.0 and \texttt{Syft} version: 1.33.0. This might not be true for future versions.}.

As discussed in Section~\ref{sec:discussion}, despite \texttt{Syft} and \texttt{Trivy} being able to identify the same dependencies, their outputs may differ slightly due to different naming conventions. In our experimental setup, we removed platform-specific information from \texttt{Ruby} dependencies to align the version conventions and scopes of both \texttt{Trivy} and \texttt{Syft}.

\subsection{Vulnerability Scan and Verification}
\texttt{Grype} is developed by the same organization as the SBOM generator \texttt{Syft} and is a widely-used vulnerability scanning tool that has been benchmarked in numerous studies~\cite{yu2024correctness,rabbi2024sbom,cofano2024sbom}. \texttt{Trivy} is a similarly popular tool capable of detecting vulnerabilities from SBOMs, and it is frequently compared with \texttt{Grype} in related research~\cite{yu2024correctness,rabbi2024sbom,cofano2024sbom}.

As shown in Fig.~\ref{fig:vulnerablity}, we first generate SBOMs in \texttt{CycloneDX} format from the source code repositories using the aforementioned tools. We then use these SBOMs to generate corresponding vulnerability reports. To validate the reported vulnerabilities, we adopt the following procedure.
If a vulnerability advisory explicitly identifies vulnerable functions, we first search the codebase for their occurrences and then leverage the default language server protocol (LSP) linters to analyze the corresponding function references. If no specific vulnerable functions are mentioned, we manually inspect the source code to determine whether the vulnerability is present. 

Given the time-consuming nature of this manual verification due to the poor vulnerability advisory metadata (more on this in Section~\ref{sec:metadata_bottleneck}), we adopt a systematic procedure to select four repositories per language while maintaining statistical rigor. First, we analyze the distribution of vulnerabilities reported by the scanners (detailed in the Appendix~\ref{sec:vuln_stats}). For each language, our goal is to select three repositories whose vulnerability counts match the median. If fewer than three repositories hold the exact median value, we choose those with counts closest to the median. In contrast, if more than three repositories share the median vulnerability count, we further evaluate the distribution of their dependencies. From this subset, we select three repositories corresponding to the 25\%, 50\%, and 75\% percentiles of dependency counts. Finally, to ensure broader coverage and mitigate selection bias, we randomly select an additional repository per language. This process results in four selected repositories per language, yielding a total of 16 repositories for manual inspection

\subsection{Data Collection}

Since the authors of \cite{yu2024correctness} do not provide the list of repositories used in their study, we follow their methodology by sourcing GitHub repositories from the corresponding \texttt{awesome-LANGUAGE} lists. These lists feature curated and popular GitHub repositories for specific programming languages. For each language, we select the \texttt{awesome-LANGUAGE} list with the highest number of stars, as this typically indicates the list's popularity and relevance.

Repositories within the same programming language may employ different PMs. To ensure the relevance of our dataset, we retain only those repositories that contain the aforementioned PM project files in the root directory. Although some repositories might place these files in subdirectories, GitHub enforces a limit of 5,000 API requests per user per hour for browsing files within folders. To fit within this quota, we restrict our parsing to the root directory exclusively. For repositories with multiple branches, we clone the default branch. If the repository owner specified no default branch, we prioritize the \texttt{main} branch, followed by the \texttt{master} branch. Repositories containing neither of the branches mentioned above are excluded from our dataset. For vulnerability verification, due to the lack of automated tools for large-scale validation, we randomly selected a code repository for each programming language. For each selected repository, we analyzed the two vulnerability reports derived from the SBOMs generated by \texttt{Grype} and \texttt{Trivy}, respectively.

\subsection{SBOM Ground Truth}
Strong PMs record all installed dependencies, including transitive dependencies, within a lock file during the installation process. This lock file precisely captures the exact set of dependencies required to build the software from a clean environment at a given point in time, thereby ensuring that identical software environments can be replicated. Since SBOMs are intended to document all direct and transitive dependencies necessary for software construction, lock files provide a reliable representation of this information. Therefore, in our experiments, we use the lock files generated by strong PMs as the ground truth. For each project file, we extract the dependencies from the corresponding lock file as a set $G$ of \texttt{(package, version)} pairs to serve as the basis for comparison.

\subsection{Comparison Metrics}

\subsubsection{Jaccard Similarity}
For each repository, \texttt{Syft} and \texttt{Trivy} generate one SBOM each, listing the identified required software components. These SBOMs are processed to extract sets of \texttt{(package, version)} pairs, representing the dependencies identified by each tool. To evaluate the similarity between the two sets generated from two different SBOM generators, respectively, denoted as $A$ and $B$, we employ the Jaccard similarity $J(A, B)$, shown in Eq.~\ref{equ:jaccard}, to quantify the overlap between the sets relative to their union.
\begin{equation}
    J(A, B) = \frac{|A \cap B|}{|A \cup B|}
    \label{equ:jaccard}
\end{equation}
A higher Jaccard similarity score indicates a greater overlap between the two sets, suggesting that both tools exhibit similar behavior when processing the same input. Conversely, a lower score reflects greater divergence, highlighting differences in how the two tools parse and generate SBOM files. Ideally, if the SBOM tools adhere to the same standards for parsing and generating SBOM files from lock files, the Jaccard similarity score would be exactly equal to 1. 

\subsubsection{Accuracy Score}
For each repository, the generated SBOMs contain the software requirements. We compare the generated requirements with the ground truth, and then we calculate the accuracy of an SBOM generator. As mentioned above, we use $A$ to denote the software dependencies extracted from an SBOM and $G$ to denote the dependencies extracted from the ground truth. The accuracy is calculated as follows:
\begin{equation}
    S(A, G) = \frac{|A \cap G|}{|G|}
    \label{equ:accuracy}
\end{equation}
A higher accuracy score indicates that the SBOM generator has successfully captured a greater proportion of the actual software dependencies, reflecting better alignment with the ground truth. The accuracy score allows us to quantitatively evaluate the performance of SBOM generators and their ability to produce reliable SBOMs.

\subsubsection{False Positive Rate}
For each vulnerability report, which contains a list of potential vulnerabilities, we manually inspect the source code repository and verify their existence. We calculate the false positive rate for each vulnerability report to measure the trustworthiness of the vulnerability scanner. We use $V_r$ to denote the total number of vulnerabilities reported, and $V_c$ to denote the number of vulnerabilities that are included in $V_r$ and confirmed in the source code. The false positive rate is calculated as follows:
\begin{equation}
    F(V_r, V_c) = \frac{V_r - V_c}{V_r}
\end{equation}
A lower false positive rate indicates that the vulnerability scanner is more precise in correctly identifying genuine security vulnerabilities, while minimizing the number of incorrectly flagged issues. In addition, a low false positive rate tool produces more actionable reports and reduces the manual effort required by developers to verify and triage alerts. Ideally, if vulnerability reports identify all vulnerabilities correctly, the false positive rate would be 0.

\section{Evaluation}\label{sec:evaluation}
In this section, we describe our dataset and evaluate the performance of SBOM tools with lock files as input. We aim to answer the following evaluation questions (EQs):

\begin{enumerate}
    \item \textbf{EQ 1:} Can we generate a lock file for the projects containing only a project file?
    \item \textbf{EQ 2:} Starting from the same lock file, will different SBOM tools generate \textbf{consistent} results from lock files?
    \item \textbf{EQ 3:} Can SBOM tools generate \textbf{accurate} results from lock files?
    \item \textbf{EQ 4:} Starting from accurate SBOMs, can vulnerability scanners generate \textbf{accurate} vulnerability reports?
\end{enumerate}

To answer \textbf{EQ1}, we first evaluate the feasibility of generating lock files for repositories that rely solely on project files. Using the resulting set of lock files from both pre-existing and newly generated, we address \textbf{EQ2} by assessing the consistency between \texttt{Syft} and \texttt{Trivy} via Jaccard similarity scores. For \textbf{EQ3}, we establish the ground truth by directly parsing the lock files and measuring the accuracy of the generated SBOMs against this baseline. Finally, regarding \textbf{EQ4}, we feed these verified, high-fidelity SBOMs into vulnerability scanners and manually verify the reported vulnerabilities against the source code to determine the false positive rate.

It is worth noting that both CycloneDX and SPDX are supported as output formats by \texttt{Syft} and \texttt{Trivy}, and they produce identical results, modulo their syntax. Therefore, the results presented in this section are equally true when using either one of those standards as input for the evaluation. Our experiments demonstrate that, even with project files only, generating lock files from project files allows SBOM tools to achieve highly accurate results. These findings highlight that utilizing lock files as input is the best practice for SBOM tools. To further illustrate this, we conduct a case study as a concrete example (Section~\ref{sec:case_study}) and discuss related threats to validity in Section~\ref{sec:threats_to_validity}.

\subsection{Dataset Description}
\begin{table}[t]
\centering
\caption{Distribution of project files and lock files in the dataset.}
\label{tab:project_distribution}


\begin{tabular}{llrrr} 
\toprule
\textbf{Language} & 
\thead{\textbf{Package} \\ \textbf{Manager}} & 
\thead{\textbf{Project File} \\ \textbf{Only}} & 
\thead{\textbf{Project File + } \\ \textbf{Lock File}} & 
\textbf{Total} \\
\midrule
Python & Poetry & 220 & 23 & 243 \\
Rust & Cargo & 552 & 481 & 1,033 \\
Ruby & Bundler & 598 & 159 & 757 \\
PHP & Composer & 303 & 78 & 381 \\
\bottomrule
\end{tabular}
\end{table}

Our dataset comprises 2,414 repositories spanning four different PMs and their associated programming languages. Table~\ref{tab:project_distribution} describes the statistics about our dataset. \textit{Project File Only} indicates that there is only the project file found in the root folder of that repository, whereas \textit{Both} indicates that both the project file and the corresponding lock file are available. It is worth noting that  \texttt{Poetry} is an unofficial PM, and Python supports several other popular PMs, including \texttt{Pip} and \texttt{Setuptools} (both weak PMs) and \texttt{Pipenv} (strong PM). Hence, the number of Python repositories using \texttt{Poetry} is relatively lower compared to other languages. In contrast, Rust uses \texttt{Cargo}, its official and sole PM, leading to the largest number of repositories depending on it. Finally, the relatively lower popularity of Ruby and PHP compared to other languages contributes to their smaller repository counts.

\subsection{Feasibility of Lock File Generation}
Since strong PMs provide native support for lock file generation, we evaluate whether lock files can be reconstructed for repositories that only contain project files but lack existing lock files. For this purpose, we develop automated scripts that invoke the corresponding PMs to generate lock files directly from project files.

As shown in Table~\ref{tab:lock_file_generation}, lock file generation is highly successful for \texttt{Rust}, where over 96\% of repositories can be resolved automatically. Manual inspection of the failed cases indicates that most failures stem from legacy dependencies that are no longer resolvable under the current package registry state (e.g., removed versions or changed package name). Similarly, due to outdated versions of \texttt{PHP} and \texttt{Ruby} being installed in our experimental environment, many software projects specify newer version constraints for \texttt{PHP} and \texttt{Ruby}, which our automated scripts could not accommodate smartly, resulting in errors. However, these issues can be resolved by manually installing the required updated versions and toolchains.

In contrast, Python presents a unique challenge, with a success rate of 19.1\%. This lower automated conversion rate highlights the fragmentation of the Python build ecosystem rather than an inability to generate lock files. Although \texttt{pyproject.toml} is the standard configuration~\cite{PEP518}, it supports diverse build backends (e.g., \texttt{setuptools}, \texttt{flit}) that are not universally compatible with a single strong PM like \texttt{Poetry}. In our experiment, we strictly employed \texttt{Poetry} to maintain consistency. As a consequence, our automated methodology compared to other languages results in a higher failure rate. For practitioners, these cases can be resolved by selecting backend-compatible tools (e.g., \texttt{uv} or \texttt{pip-tools}), as we discuss in Section~\ref{sec:discussion}. 

\begin{table}[tb]
\centering
\caption{Lock file generation across different programming languages.}
\label{tab:lock_file_generation}
\begin{tabular}{lccc}
\toprule
\textbf{Language} & \textbf{Successful} & \textbf{Failed} & \textbf{Success Rate} \\
\midrule
Python      & 42  & 178 & 19.1\% \\
Rust        & 532 & 20  & 96.4\% \\
Ruby        & 456 & 142 & 76.3\% \\
PHP         & 165 & 138 & 54.5\% \\
\bottomrule
\end{tabular}
\end{table}

It is worth noting that for repositories with pre-existing lock files, we also verified the capability to regenerate them from project files to ensure dependency freshness. We apply the same automation pipeline to these repositories. The total number of usable lock files per language is reported in Table~\ref{tab:merged_sbom_results}. Using the resulting lock files, we generate SBOMs with \texttt{Syft} and \texttt{Trivy} for downstream analysis.

\begin{tcolorbox}[
    colback=gray!10, 
    colframe=black,  
    sharp corners=northwest, 
    rounded corners=all, 
    width=\linewidth, 
    boxrule=0.8pt, 
    arc=6pt, 
    left=2mm, 
    right=2mm, 
    ]
\textbf{Answer to EQ 1:} Lock files can be retroactively generated for legacy repositories to enable accurate SBOM creation. Our results show that for integrated ecosystems like Rust, this process can be done automatically, in most cases. While fragmented ecosystems like Python require selecting appropriate build tools to handle diverse backends, the transition from project files to lock files is practically achievable and serves as a necessary prerequisite for SSC security.
\end{tcolorbox}

\subsection{Consistency Across SBOM Tools}
To evaluate the consistency of SBOM generation, we use the existing lock files harvested from our dataset as the primary input for both \texttt{Syft} and \texttt{Trivy}. We then compute the Jaccard similarity coefficient to quantify the agreement between the dependency sets identified by each tool.

The results are definitive: we observed a \textbf{perfect Jaccard similarity score of 1.0} with \textbf{zero variance} across all tested repositories in \texttt{Python}, \texttt{Rust}, \texttt{Ruby}, and \texttt{PHP}. This stands in clear contrast to prior studies~\cite{yu2024correctness, o2023impacts} that reported significant discrepancies when relying on project files. Our findings demonstrate, thanks to a large scale evaluation,  that lock files effectively eliminate the ambiguity associated with dependency resolution. Consequently, when provided with this deterministic input, different SBOM generators are capable of producing identical, high-fidelity outputs. Based on this evidence, we answer \textbf{EQ2} as follows:

\begin{tcolorbox}[
    colback=gray!10, 
    colframe=black,  
    sharp corners=northwest, 
    rounded corners=all, 
    width=\linewidth, 
    boxrule=0.8pt, 
    arc=6pt, 
    left=2mm, 
    right=2mm, 
    ]
\textbf{Answer to EQ 2: } By using lock files, we achieve perfect consistency between different SBOM tools across all studied languages, indicating that lock files are a reliable input source for generating consistent and reproducible SBOM files.
\end{tcolorbox}

\subsection{Accuracy against Ground Truth}
The consistency between \texttt{Syft} and \texttt{Trivy} alone does not confirm the reliability of the SBOMs they generate, as it fails to indicate whether both tools produce accurate or inaccurate results. Consequently, we validate the generated SBOMs against a ground truth, which is obtained from the lock files, to determine their correctness. Since \texttt{Syft} and \texttt{Trivy} produce the same result when using each file, their results are grouped together in a single column for comparison. 
Table~\ref{tab:merged_sbom_results} presents the experimental results. The \textit{Trivy \& Syft} column lists dependencies uniquely identified by these tools, while the \textit{Ground Truth} column shows dependencies present in the ground truth but missing from the SBOMs. The \textit{Overlap} column indicates dependencies found in both the SBOMs and the ground truth, and the \textit{Accuracy} column is derived from these values. From the results, we conclude our answer to \textbf{EQ3}.

\begin{tcolorbox}[
    colback=gray!10, 
    colframe=black,  
    sharp corners=northwest, 
    rounded corners=all, 
    width=\linewidth, 
    boxrule=0.8pt, 
    arc=6pt, 
    left=2mm, 
    right=2mm, 
    ]
\textbf{Answer to EQ 3: } The outputs of \texttt{Syft} and \texttt{Trivy} are fully consistent with the ground truth, indicating that SBOM generators are able to produce complete and accurate SBOMs, if provided with lock files.
\end{tcolorbox}

It is important to note that this study focuses exclusively on the correctness of the \texttt{name} and \texttt{version} of each component. As discussed in Section~\ref{sec:background}, SBOMs also encompass additional details such as \texttt{license}, \texttt{manufacturer}, and other metadata. Since lock files do not include this information, generating a complete and accurate SBOM requires more than simply referencing the lock file.

\begin{table}[tb]
\centering
\caption{Statistics of generated lock files and SBOM accuracy evaluation results against the ground truth.}
\label{tab:merged_sbom_results}
\resizebox{\columnwidth}{!}{%
\begin{tabular}{lccccc}
\toprule
\multirow{2}{*}{\textbf{Language}} & \multirow{2}{*}{\begin{tabular}[c]{@{}c@{}}\textbf{No. of}\\ \textbf{Lock Files}\end{tabular}} & \multicolumn{3}{c}{\textbf{Dependency Comparison}} & \multirow{2}{*}{\textbf{Accuracy}} \\
\cmidrule(lr){3-5}
 &  & \begin{tabular}[c]{@{}c@{}}\textbf{Trivy \& Syft}\\ \textbf{Unique}\end{tabular} & \begin{tabular}[c]{@{}c@{}}\textbf{Ground Truth}\\ \textbf{Unique}\end{tabular} & \begin{tabular}[c]{@{}c@{}}\textbf{No. Overlapped}\\ \textbf{Dependencies}\end{tabular} &  \\
\midrule
Python      & 65    & 0 & 0 & 1,997   & 100.00\% \\
Rust        & 1,013 & 0 & 0 & 237,576 & 100.00\% \\
Ruby        & 615   & 0 & 0 & 29,420  & 100.00\% \\
PHP         & 243   & 0 & 0 & 2,683   & 100.00\% \\
\bottomrule
\end{tabular}
}
\vspace{-3mm}
\end{table}

\subsection{Vulnerability Report Evaluation}\label{sec:manual_vulnerablity}

Using the SBOMs from the previous experiment, we evaluate the performance of two popular SBOM-based vulnerability scanners: \texttt{Trivy} and \texttt{Grype}. To adhere to the recommended workflow for each tool, we generated two distinct SBOMs from the lock file of each code repository. The first SBOM was generated with \texttt{Syft} and subsequently scanned by its companion tool, \texttt{Grype}. The second SBOM was generated and scanned using \texttt{Trivy}. This paired approach ensures that each scanner operates on an SBOM created by its corresponding generation tool, aligning with the developer's recommendations.

\begin{table}[tb]
\centering
\caption{Comparison of vulnerabilities detected by Grype and Trivy.}
\resizebox{\columnwidth}{!}{%
\begin{tabular}{l rr r rr r}
\toprule
& \multicolumn{3}{c}{\textbf{Grype}} & \multicolumn{3}{c}{\textbf{Trivy}} \\
\cmidrule(lr){2-4} \cmidrule(lr){5-7}
\textbf{Language} & \textbf{No. Repos} & \textbf{No. Vulns} & \textbf{Avg.} & \textbf{No. Repos} & \textbf{No. Vulns} & \textbf{Avg.} \\
\midrule
Python & 13   & 148   & 11.38 & 8    & 109   & 13.62 \\
Rust   & 348  & 2864  & 8.23  & 347  & 2858  & 8.24  \\
Ruby   & 124  & 2562  & 20.66 & 120  & 2548  & 21.23 \\
PHP    & 11   & 41    & 3.73  & 11   & 40    & 3.64  \\
\midrule
\textbf{Total} & \textbf{496} & \textbf{5615} & \textbf{11.32} & \textbf{486} & \textbf{5555} & \textbf{11.43} \\
\bottomrule
\end{tabular}%
}
\label{tab:vuln_resized}
\vspace{-3mm}
\end{table}

Initially, we scanned all our code repositories for vulnerabilities using both SBOM-based scanners, and the results are summarized in Table~\ref{tab:vuln_resized}. Some repositories do not rely on any third-party libraries and therefore cannot yield vulnerability findings. Consequently, the \textit{No.~Repos} column reports only the number of repositories in which at least one vulnerability was identified, while the \textit{No.~Vulns} column gives the total count of vulnerabilities detected across them. On average, vulnerable repositories contain more than three vulnerabilities each, with \texttt{Ruby} projects showing the highest density at over 20 per repository.

Although the large number of vulnerabilities per vulnerable repository might raise a significant and widespread security concern, they do not distinguish between true and false positives and, therefore, cannot represent the practical trustworthiness of the scanners. To further investigate this, we conduct a follow-up validation experiment. However, there is an absence of automated source code analysis tools for our target languages, unlike specialized tools such as~\citet{zhao2024covsbom} for \texttt{Java}. Hence, we transition from our previous large-scale analysis to a deep-dive manual validation, which is inherently labor-intensive. Therefore, we prioritize a deep-dive analysis on a small and representative set of repositories over a broader but superficial statistical count. Accordingly, we randomly select one repository that contains fewer than 20 vulnerabilities reported for each language, scan it with both scanners, \texttt{Grype} and \texttt{Trivy}, and then manually inspect the source code to validate each reported vulnerability.

\begin{table}[tb]
\centering
\caption{Reported vs. confirmed vulnerabilities. $V_c$ is confirmed if the code \textbf{exists} in the repository. $V_r$: reported, $V_c$: confirmed, FPR: false positive rate.}
\label{tab:vulnerability_validation}
\resizebox{\linewidth}{!}{%
\begin{tabular}{llrrr}
\toprule
Language & Repository & $V_r$ & $V_c$ & FPR \\
\midrule
Rust & \texttt{guoxbin/dtool} & 15 & 1 & 93.3\% \\
     & \texttt{0xlane/pe-sign} & 4 & 0 & 100.0\% \\
     & \texttt{sagiegurari/duckscript} & 4 & 1 & 75.0\% \\
     & \texttt{serverlesstechnology/cqrs} & 4 & 0 & 100.0\% \\
\midrule
Python & \texttt{mingrammer/diagrams} & 3 & 0 & 100.0\% \\
       & \texttt{beetbox/beets} & 9 & 4 & 55.6\% \\
       & \texttt{sdispater/pendulum} & 7 & 0 & 100.0\% \\
       & \texttt{pyca/cryptography} & 9 & 0 & 100.0\% \\
\midrule
PHP & \texttt{jolicode/JoliCi} & 8 & 0 & 100.0\% \\
    & \texttt{phpbrew/phpbrew} & 3 & 0 & 100.0\% \\
    & \texttt{FriendsOfPHP/pickle} & 3 & 0 & 100.0\% \\
    & \texttt{getgrav/grav} & 4 & 0 & 100.0\% \\
\midrule
Ruby & \texttt{abhidsm/time\_diff} & 15 & 0 & 100.0\% \\
     & \texttt{honeybadger-io/incoming} & 9 & 0 & 100.0\% \\
     & \texttt{camping/camping} & 8 & 3 & 62.5\% \\
     & \texttt{grosser/fast\_gettext} & 8 & 0 & 100.0\% \\
\bottomrule
\textbf{Total} & -- & \textbf{113} & \textbf{9} & \textbf{92.0\%} \\
$Total_{cov}$ & -- & \textbf{42} & \textbf{9} & \textbf{78.6\%} \\
\bottomrule
\end{tabular}%
}
\end{table}

Table~\ref{tab:vulnerability_validation} presents the results of our manual validation, which revealed an exceptionally high false positive rate. The $V_r$ in Table~\ref{tab:vulnerability_validation} represents the union of findings from both \texttt{Grype} and \texttt{Trivy}. It is important to note that, starting from the same accurate SBOM, we observed a near-perfect agreement between the two scanners. Despite the fact that these tools utilize distinct vulnerability databases. From the 113 potential vulnerabilities reported across four repositories, our in-depth manual inspection confirmed only 9 true positives\footnote{Following ethical security research practices, we have responsibly disclosed these findings to the respective maintainers. Several authors have already acknowledged and addressed these issues.}. This corresponds to a staggering overall false positive rate of 92.0\%. This result demonstrates a pronounced discrepancy between automated reports and actual security risks. Unfortunately, facing such a high false positive rate, we cannot help questioning the usability and trustworthiness of their methods. 

We notice that a substantial portion of the vulnerabilities detected are not from developer-introduced third-party dependencies, but from the foundational runtime libraries of the programming languages themselves. For example, the scanners flagged ten vulnerabilities within the \texttt{activesupport@3.0.1} library. As a core dependency of \texttt{Ruby on Rails}, \texttt{activesupport} is a comprehensive library offering a vast collection of functionalities. The vulnerability reports almost exclusively cited issues related to HTTP-based attacks. However, this repository only invoked a very small and isolated subset of \texttt{activesupport}'s capabilities. A manual inspection confirmed that the vulnerable modules were never imported or referred, rendering all ten reported vulnerabilities untriggerable and, therefore, not applicable to the security posture of this specific application. 

Our evaluation strongly indicates a systemic limitation in the current generation of SBOM scanners rather than an implementation defect. The fundamental issue is that these tools primarily operate at the package level, verifying the presence of vulnerable library versions listed in dependency manifests without analyzing the actual function call. To validate this hypothesis, we conducted a further experiment by analyzing whether the specific vulnerable functions cited in the reports were ever actually used by the code. The results, presented in the $Total_{cov}$ row of Table~\ref{tab:vulnerability_validation}, show that the coverage-based analysis successfully eliminated 61.9\% reported false vulnerabilities, reducing the number of alerts requiring manual inspection from 113 to 42. It demonstrates that a significant portion of false positives can be systematically filtered out. This lack of context awareness generates a significant volume of false positives, leading to substantial alert fatigue. As developers become inundated with alerts, they must manually check each one, a process that not only misdirects engineering efforts but also fosters an environment where critical, exploitable vulnerabilities are more likely to be overlooked in tons of false alarms. As a consequence, paradoxically, this noisy security approach can ultimately harm software security. The effectiveness of our call analysis strongly suggests the value of developing coverage-based SBOM vulnerability scanners for SSC security ecosystems.

\begin{tcolorbox}[
    colback=gray!10, 
    colframe=black,  
    sharp corners=northwest, 
    rounded corners=all, 
    width=\linewidth, 
    boxrule=0.8pt, 
    arc=6pt, 
    left=2mm, 
    right=2mm, 
    ]
\textbf{Answer to EQ 4: } An accurate SBOM is a necessary but not a sufficient foundation for vulnerability management. It confirms what components are present, but not how they are used. Consequently, scanners relying only on SBOMs will inevitably flag unreachable vulnerabilities. To bridge this gap and reflect a software's true risk posture, adopting a coverage-based analysis to determine code reachability is essential.
\end{tcolorbox}

\subsection{Case Study: Transition from Weak to Strong PM}\label{sec:case_study}

We present a concrete real-world case demonstrating how to generate an accurate SBOM using a lock file from a repository managed by a weak PM. Specifically, we analyze the \texttt{gym} repository from \texttt{OpenAI}'s GitHub repository. \texttt{gym} is a widely used Python library for reinforcement learning, with over 1,315 dependent packages and 60,881 third-party projects relying on it, making it one of the most popular reinforcement learning libraries~\cite{farama2022announcement}. However, prioritizing native compatibility, \texttt{gym} employs the weak PM \texttt{pip}, the official PM of \texttt{Python}, and uses \texttt{requirements.txt} to manage its dependencies. As discussed earlier, this project file does not capture indirect dependencies, potentially leading SBOM tools to produce inaccurate results.

We use the \texttt{requirements.txt} file as the input for both of the previously mentioned SBOM generators. In the \texttt{requirements.txt} file, 15 dependencies are explicitly listed. However, both \texttt{Syft} and \texttt{Trivy} detect only 5 dependencies. Further investigation reveals that these detected dependencies are constrained strictly by the \texttt{==} operator, while dependencies specified with a version range (e.g., \texttt{>=}) are not identified. Moreover, a discrepancy is observed between the two tools. In the \texttt{requirements.txt} file, the dependency \texttt{typing\_extensions} is altered to \texttt{typing-extensions} by \texttt{Trivy}, whereas \texttt{Syft} retains the original format.

Following the workflow mentioned above, we parse the \texttt{require
ments.txt} file and transfer the dependencies, along with their corresponding version constraints, into the \texttt{pyproject.toml} file, which is the project file for the strong PM \texttt{Poetry}. Subsequently, we generate the corresponding lock file and use both \texttt{Syft} and \texttt{Trivy} to construct an SBOM based on this lock file and compare the results. In total, 32 dependencies are now identified consistently. Based on the newly generated SBOM file, we conduct a vulnerability scan and no vulnerabilities are found in this repository. 

This use case shows how easily we can generate an accurate SBOM and conduct a vulnerability scan even when a certain project uses a weak PM and project file to manage its dependencies. Our experiment demonstrates that bridging it to a strong PM flow (as detailed in our migration guidelines in Section~\ref{sec:migrate}) effectively eliminates the noise caused by incomplete dependency resolution. In this case study, SBOM tools produce identical and accurate reports that match the ground truth, and the subsequent vulnerability scan validates the security of the repository, with no vulnerabilities detected.

\subsection{Feasibility of Automated Reachability Analysis}\label{sec:go_automation}

Our manual verification in Section~\ref{sec:manual_vulnerablity} exposes a critical issue that existing scanners produce excessive false positives. Scaling this verification to a larger dataset, however, is restricted by a data limitation. Most vulnerability advisories (e.g., CVE, GitHub Advisory) lack machine-readable \textit{affected symbol} metadata, making manual inspection the only viable option.

We use the Go ecosystem to validate the feasibility of automating this process. Unlike others, the Go vulnerability database explicitly provides structured symbol data. Leveraging this, we implement our Proof-of-Concept (PoC) automated tool\footref{note4} to perform reachability analysis following the methods mentioned before and evaluate it on 12 popular Go projects. We show that even a lightweight, conservative call reachability analysis can eliminate a substantial fraction of evidently irrelevant vulnerability reports.

Table~\ref{tab:go_summary_aggregate} summarizes the results (We demonstrate the detailed breakdown in Appendix A). Out of 121 reported vulnerabilities, our automated analysis confirm that only 31 were present, representing a \textbf{74.4\% reduction} in false positives. The improvement is particularly significant for complex projects. For instance, we reduce 90\% reported vulnerabilities in \texttt{Prometheus}. We conclude that if upstream advisories provide granular metadata, reachability analysis can be fully automated beyond \texttt{Go} ecosystem. Furthermore, we can significantly reduce the false positive rate and alleviate the developers' alert fatigue.

\begin{table}[tb]
\centering
\caption{Aggregate results of automated reachability analysis across 12 Go projects. (Detailed breakdown in Appendix A)}
\vspace{-3mm}
\label{tab:go_summary_aggregate}
\resizebox{\columnwidth}{!}{%
\begin{tabular}{cccc}
\toprule
\textbf{Projects} & \textbf{Total Reported} & \textbf{Confirmed Reachable} & \textbf{Reduction Rate} \\
\midrule
12 & 121 & 31 & \textbf{74.4\%} \\
\bottomrule
\end{tabular}
}

\end{table}

\section{Discussion}\label{sec:discussion}

In this section, we first discuss the threats to validity of our study. We then introduce the reasons behind the differences observed in our SBOM evaluation and how to transition from weak PMs. Subsequently, we outline best practices to enhance the security of SSC and the reliability of SBOM tools.

\subsection{Threats to Validity}\label{sec:threats_to_validity}

For the SBOM experiment, we construct our ground truth based on the lock files generated by strong PMs. However, our experimental setup is limited to a fixed environment. It is possible that these PMs produce inconsistent dependencies across different environments, such as varying operating systems, CPU architectures, and other system configurations. SBOMs tailored to specific runtime environments may offer greater accuracy, which can be a direction for future exploration. 
Regarding vulnerability verification, our manual inspection process is subject to human error. It is possible that certain vulnerabilities were inadvertently overlooked or that some could be exploited through novel or complex attack vectors beyond the scope of our analysis. Our manual verification is limited to 4 repositories due to the labor-intensive nature of deep code auditing and function call tracing. However, the root causes of false positives proved remarkably consistent across diverse languages. Consequently, extending the manual effort would yield diminishing returns. We prioritized a granular, deep-dive analysis over a broader but superficial statistical count to ensure high-fidelity insights
Our study focuses solely on static analysis under normal software development conditions. Consequently, techniques such as obfuscation, reflection, dynamic code loading, and dependency vendoring~\cite{birsan2021dependency_confusion,miller2023we_feel,qu2017dydroid,rscox_vgo} are not addressed, which may limit the generalizability of our results.

\subsection{Factors Contributing to SBOM Discrepancies}\label{sec:discrepancy}

\subsubsection{Undefined Behavior}
Although SBOM standards are publicly available, the implementation of SBOM tools may vary. Differences in how those tools handle the input files can result in inconsistent outputs. For instance, when generating SBOMs from \texttt{poetry.lock}, we observe that if a package in the lock file has the attribute \texttt{category} set to \texttt{dev}, early versions of \texttt{Trivy} (before \texttt{v0.59.1}) omit the package, while \texttt{Syft} includes it in the SBOM. While the latest versions mitigate this issue, a systematic solution requires standardized behavior.

\subsubsection{Incomplete Dependency Resolve}
SBOM tools may fail to resolve all dependencies, leading to incomplete SBOMs. For example, when generating SBOMs from \texttt{requirements.txt} files, both \texttt{Trivy} and \texttt{Syft} only resolved direct dependencies explicitly specified in the \texttt{package==version}-like format, resulting in incomplete and inconsistent SBOMs. This issue can be mitigated by using lock files, which comprehensively record all dependencies, including transitive ones, and specify exact versions. This ensures that the generated SBOMs are both complete and accurate.

\subsubsection{Inconsistent Naming Convention}
Dependency naming conventions may differ across PMs, leading to inconsistencies in the generated SBOMs. For example, when generating SBOMs from \texttt{Gemfile.lock}, we find discrepancies in how \texttt{Trivy} and \texttt{Syft} name dependencies. Ruby's \texttt{Gemfile.lock} may include dependencies with platform-specific annotations, such as \texttt{sqlite3 (2.0.2-aarch
64-linux-gnu)}. \texttt{Trivy} outputs only the name and version, omitting the architecture, while \texttt{Syft} appends the architecture after the version as part of the full version name. These differences result in inconsistent dependency representations, which can affect subsequent analyses. Standardizing dependency naming across PMs can address this issue and ensure consistent SBOM generation.

\subsubsection{Directory Scanning Behavior}

When provided with a directory as input, SBOM tools often scan the entire directory recursively to identify project files for analysis. However, discrepancies in scanning behavior can lead to inconsistent results. For instance, some tools include auxiliary folders such as \texttt{.ci}, \texttt{test}, or \texttt{docker}, in addition to the \texttt{GitHub Actions} folder, while others exclude them. These folders are not the software's direct dependencies but are used for CI/CD pipelines or release processes. The inconsistency in whether these auxiliary folders are included during scanning contributes to differences in the generated SBOMs~\cite{balliu2023challenges}.

Additionally, other factors may exacerbate inconsistencies in SBOM generation. For instance, implementation errors in the tools themselves, such as incorrect handling of file paths or misinterpretation of configuration files, can introduce inaccuracies. Similarly, issues with parsing lock files, such as failing to recognize all dependencies or incorrectly resolving transitive dependencies, can further affect the accuracy of the SBOMs. These challenges highlight the need for standardization in the way SBOM tools scan directories and interpret project metadata to ensure reliable and consistent results.

\subsection{Transition From Weak to Strong PM}~\label{sec:migrate}
Our results clearly indicate that migrating to strong PMs, that produce lock files, is imperative for accurate SBOM generation and subsequent security analysis. We propose a practical migration guideline addressing two common scenarios found in legacy codebases. For Python, the fragmented ecosystem and the prevalence of its default weak PM necessitate manual effort for developers to migrate toward strong alternatives. In contrast, ecosystems such as Rust, Ruby, and PHP benefit from official package managers that function as strong PMs by default. In these cases, the primary barrier to robust SBOM generation is user adoption rather than a lack of tooling capability.

\subsubsection{Bridging the Gap for Weak PMs.}
For projects adopting weak PMs (e.g., Python's \texttt{pip} with \texttt{requirements.txt}), completely refactoring to a strong PM format (e.g., \texttt{Poetry}'s \texttt{pyproject.toml}) can be labor-intensive. To mitigate this migration cost, we recommend utilizing modern bridge tools that offer interoperability. For example, the emerging tool \texttt{uv}~\cite{UV_Repo} allows developers to treat \texttt{requirements.txt} as a source manifest and employs a \texttt{pip-compile} to resolve dependencies deterministically. This process generates a lock file that captures the exact dependency tree, enabling the generation of accurate SBOMs without forcing a disruptive overhaul of the existing project structure.
\subsubsection{Producing Lock Files in Strong PMs.}
For repositories already utilizing strong PMs (e.g., Rust's \texttt{Cargo}) but failing to track lock files in version control, the solution is procedural rather than technical. Developers must integrate lock file generation into the build pipeline and commit these files as the source of truth. Modern strong PMs facilitate this efficiently; for instance, \texttt{Cargo} supports a \textit{dry-run} mechanism that resolves the dependency graph and updates the \texttt{Cargo.lock} file without triggering the resource-intensive download and installation of artifacts. This ensures high-fidelity SBOM generation with minimal overhead.

Ultimately, facilitating this transition is not merely an operational convenience but a security necessity. By standardizing on lock-file-based generation through these migration paths, we establish the high-fidelity ground truth required to eliminate upstream noise, thereby significantly enhancing the reliability of the entire downstream vulnerability management lifecycle.

\subsection{The Bottleneck of Unstructured Vulnerability Metadata}
\label{sec:metadata_bottleneck}
Our automated analysis in the Go ecosystem shows that automated reachability verification is feasible when the vulnerability metadata is available, reducing considerable false positives. However, unlike the \texttt{Go} ecosystem, for other languages, information about affected functions is typically buried in unstructured natural language text. Consequently, automating reachability analysis for these languages without high-fidelity, machine-readable metadata is highly impractical to execute reliably at scale, thereby restricting us to deep, time-consuming, manual validation. In fact, we have directly discussed this limitation with the developers of major SBOM tools. They confirmed that the absence of function-level metadata in mainstream vulnerability advisories (e.g. NVD, GHSA) is a known, critical bottleneck, but extracting this information at scale remains an unsolved industry challenge. This consensus further justifies our deep manual validation as the only rigorous approach currently available.

Based on this limitation, we call for a community effort to improve mainstream vulnerability advisories by including standardized, machine-readable \texttt{affected\_symbols} in structured form, rather than only in narrative text, which would significantly improve the precision and scalability of reachability-aware analysis and make vulnerability reports more actionable in practice.

\subsection{A Practical Pipeline for Actionable Vulnerability Reporting}
Based on our findings, we outline a practical, two-stage pipeline designed to guide practitioners from the current state of noisy, package-level alerts to more actionable, context-aware vulnerability reports. This pipeline defines clear responsibilities for both software developers and security tool builders~\footnote{We have contacted the developers of both \texttt{Trivy} and \texttt{Grype} regarding these issues, and are currently in discussions with them.}.

\subsubsection{Stage 1: Ground-Truth SBOM Generation via Lock Files}
The foundational stage of this pipeline is to ensure the integrity of the input: the SBOM itself. Our research confirms that inconsistencies arise when SBOM generators attempt to resolve dependencies from project files. To build a source of truth, this stage requires two actions:

\textbf{For Developers:} Standardize on using strong PMs that generate lock files. These lock files must be committed to version control as the definitive record of all direct and transitive dependencies for a reproducible build.

\textbf{For SBOM Tool Builders:} Treat lock files as the only authoritative source. Tools should parse lock files directly, rather than implementing complex and error-prone dependency resolution logic. This guarantees the generated SBOM is a high-fidelity representation of what is actually used in production.
\subsubsection{Stage 2: Finer Granularity Analysis via Reachability}

\textbf{For Security Scanners:} The next generation of vulnerability scanners must move beyond simple version matching and incorporate reachability analysis. As validated by our function call experiment, this layer of analysis filters out the vast majority of non-exploitable vulnerabilities by determining if the vulnerable code is actually invoked. This is the key to transforming noisy alert streams into actionable intelligence and directly combating developer alert fatigue.

\section{Conclusions and Future Work}\label{sec:conclusion}

In this study, we first address the foundational challenge of SBOM accuracy. Through a large-scale evaluation of 2,414 open-source projects, we demonstrated that utilizing lock files as the primary input enables popular SBOM generators like \texttt{Syft} and \texttt{Trivy} to produce consistent, complete, and reliable SBOMs. This establishes a crucial first step for any practical SBOM-based SSC pipeline.

However, our study reveals that even these accurate SBOMs are not a silver bullet for vulnerability management. Our manual evaluation of downstream scanners uncovered a more profound flaw in the current ecosystem: an overwhelming 92.0\% false positive rate in our sampled dataset. This finding demonstrates that the current generation of SBOM-based scanners is fundamentally inadequate for practical use, burying developers in unactionable alerts and undermining the trustworthiness of the entire process.

This limitation arises from the coarse granularity of information typically provided by SBOMs, which operate at the package level. A package may include a large number of functions, and the presence of a single vulnerable function can result in the entire package—and by extension, the software—being flagged as vulnerable, even if the function is never invoked. For practitioners, this overgeneralization directly leads to wasted resources on manual triage and encourages ignoring security alerts, which ultimately degrades the security posture.

Our work not only identifies this critical challenge but also validates a clear path forward. Our large-scale, multi-language study provides the first comprehensive, empirical evidence that reachability analysis is a fundamental requirement across the board. Our function call analysis serves as a generalizable proof-of-concept, demonstrating that reachability analysis can successfully prune a majority of these false positives. Our future work will build upon our validated approach by exploring more advanced static or dynamic analyses to handle complex cases. Solving this issue is paramount to unlocking the full potential of SBOMs and finally delivering on their promise of precise and actionable security insights for practitioners.

\bibliographystyle{ACM-Reference-Format}
\balance
\bibliography{biblio}

\appendix

\section{Detailed Results of Automated Go Analysis}

\label{app:go_results}
\begin{table*}[t]
\centering
\caption{Complete evaluation results of automated vulnerability pruning on 12 Go repositories with repository sources. ($V_r$: Reported, $V_c$: Confirmed)}
\label{tab:go_full_results}
\resizebox{2\columnwidth}{!}{%
\begin{tabular}{lllccc}
\toprule
\textbf{Project} & \textbf{Version} & \textbf{Repository URL} & \textbf{$V_r$} & \textbf{$V_c$} & \textbf{Reduction Rate} \\
\midrule
Prometheus & v3.1.0 & \texttt{https://github.com/prometheus/prometheus} & 10 & 1 & 90.0\% \\
Etcd & v3.5.18 & \texttt{https://github.com/etcd-io/etcd} & 22 & 3 & 86.4\% \\
Hugo & v0.141.0 & \texttt{https://github.com/gohugoio/hugo} & 10 & 2 & 80.0\% \\
MinIO & RELEASE.2025-01-18T00-31-37Z & \texttt{https://github.com/minio/minio} & 12 & 3 & 75.0\% \\
Rclone & v1.69.0 & \texttt{https://github.com/rclone/rclone} & 14 & 4 & 71.4\% \\
Traefik & v2.11.17 & \texttt{https://github.com/traefik/traefik} & 17 & 5 & 70.6\% \\
Lazygit & v0.45.0 & \texttt{https://github.com/jesseduffield/lazygit} & 6 & 2 & 66.7\% \\
Gitea & v1.23.0 & \texttt{https://github.com/go-gitea/gitea} & 14 & 5 & 64.3\% \\
Caddy & v2.9.1 & \texttt{https://github.com/caddyserver/caddy} & 12 & 6 & 50.0\% \\
Fzf & v0.58.0 & \texttt{https://github.com/junegunn/fzf} & 0 & 0 & - \\
Dive & v0.13.0 & \texttt{https://github.com/wagoodman/dive} & 1 & 0 & 100\% \\
Nats-server & v2.10.25-RC.1 & \texttt{https://github.com/nats-io/nats-server} & 3 & 0 & 100\% \\
\midrule
\textbf{Total} & \textbf{-} & \textbf{-} & \textbf{121} & \textbf{31} & \textbf{74.4\% (Avg)} \\
\bottomrule
\end{tabular}%
}
\end{table*}

In this appendix, we provide the comprehensive data obtained from our automated reachability analysis on the Go ecosystem, as discussed in Section \ref{sec:go_automation}. We selected 12 prominent, open-source Go repositories covering various domains, including web servers, databases, and CLI tools. To simulate a realistic scenario of software maintenance where vulnerabilities may have accumulated, we targeted historical release versions (the first released version in 2025) rather than the latest \texttt{HEAD} commits. 

Table \ref{tab:go_full_results} details the specific versions analyzed and the comparison results. The column $V_r$ (Reported) represents the number of vulnerabilities flagged by standard SBOM-based scanners based solely on package version matching. The column $V_c$ represents the number of vulnerabilities where the specific \texttt{affected\_symbols} (as defined in the Go Vulnerability Database) were confirmed to be present in the compiled binary's symbol table by our automated tool. The \textit{Reduction Rate} is calculated as $(V_r - V_c) / V_r$, indicating the percentage of false positives pruned by our reachability analysis.

The results demonstrate a consistent pattern across different types of applications: the majority of dependencies flagged as vulnerable are dead code that is eliminated during the compilation process or never invoked by the application logic. The significant reduction in false positives (74.4\% on average) empirically validates that metadata-driven reachability analysis is a highly effective strategy for improving the signal-to-noise ratio in vulnerability reporting.

Furthermore, to demonstrate the proactive utility of our approach on actively developed software, we also applied our automated analysis to the latest \texttt{HEAD} commits of these same 12 repositories. This scan identified a new, reachable vulnerability in one of the projects. Following ethical security research practices, we performed a responsible disclosure by privately reporting our findings to the project's maintainers. The vulnerability was subsequently acknowledged and patched by the developers. This real-world validation not only confirms the effectiveness of our metadata-driven reachability analysis but also underscores its value as a practical tool for developers to proactively secure their software, moving beyond simply triaging scanner alerts.

\section{Vulnerability Distribution}~\label{sec:vuln_stats}

\begin{figure}[htbp]
    \centering
    \input{figure/vulnerability_boxplot.pgf}
    \caption{Distribution of detected vulnerabilities per repository across different programming languages, comparing Trivy and Grype scanners.}
    \Description{Distribution of detected vulnerabilities per repository across different programming languages, comparing Trivy and Grype scanners.}
    \label{fig:vuln_distribution}
\end{figure}
As illustrated in Figure \ref{fig:vuln_distribution}, the volume of reported vulnerabilities varies significantly depending on the language ecosystem. Ruby repositories exhibit the highest median number of vulnerabilities and the widest variance, indicating a potentially more complex dependency chain or a higher historical vulnerability density within its ecosystem. Conversely, PHP repositories demonstrate the lowest overall vulnerability counts. However, reporting a higher raw count of vulnerabilities does not necessarily equate to better security posture. An inflated detection rate often correlates with a higher FPR, which increases alert fatigue for developers. 

\end{document}